\documentclass[%
 reprint,
preprintnumbers,
nofootinbib,
amsmath,amssymb,aps,superscriptaddress
]{revtex4-2}

\usepackage[dvips]{graphicx}
\usepackage{dcolumn}
\usepackage{bm}
\usepackage[hidelinks]{hyperref}
\usepackage[dvipsnames]{xcolor}
\usepackage[normalem]{ulem}
\usepackage{cancel}

\def\d{\mathrm{d}}

\def\O{\mathcal{O}}

\def\e{\mathrm{e}}
\def\P{\mathcal{P}}

\usepackage{soul}

\def\GeV{{\rm GeV}}
\def\MeV{{\rm MeV}}
\def\eV{{\rm eV}}

\newcommand{\rom}[1]{\uppercase\expandafter{\romannumeral #1\relax}}
\newcommand{\overbar}[1]{\mkern 1.5mu\overline{\mkern-1.5mu#1\mkern-1.5mu}\mkern 1.5mu}

\def\bal#1\eal{\begin{align}#1\end{align}}

\begin{document}

\preprint{KCL-PH-TH-2024-46}
\preprint{CTPU-PTC-24-28}

\title{
Primordial black holes from an interrupted phase transition 
}

\author{Wen-Yuan Ai}
\email{wenyuan.ai@kcl.ac.uk}

\author{Lucien Heurtier}
\email{lucien.heurtier@kcl.ac.uk}
\affiliation{Theoretical Particle Physics and Cosmology,\\ King’s College London, Strand, London WC2R 2LS, UK 
}

\author{Tae Hyun Jung}
\email{thjung0720@gmail.com}
\affiliation{Particle Theory and Cosmology Group, Center for Theoretical Physics of the Universe,\\
Institute for Basic Science (IBS), Daejeon, 34126, Korea}


\begin{abstract}

We propose a new mechanism of primordial black hole formation 
via an interrupted phase transition during the early matter-dominated stage of reheating after inflation. 
In reheating, induced by the decay of a pressureless fluid dominating the Universe at the end of inflation, dubbed as reheaton, the temperature of the radiation bath typically increases, reaching a maximum temperature $T_{\rm max}$, and then decreases. We consider a first-order phase transition induced by the increase of the temperature that is aborted as $T_{\rm max}$ is higher than the critical temperature but not sufficiently high for the bubble nucleation rate to overcome the expansion of the Universe. Although bubbles never fully occupy the space, some may be nucleated and expand until the temperature once again decreases to the critical temperature. We argue that these bubbles shrink and disappear as the temperature drops further, leaving behind macroscopic spherical regions with positive density perturbations. These perturbed regions accrete the surrounding matter (reheatons) and eventually collapse into primordial black holes whose mass continues to grow until the onset of radiation domination. 
We estimate the abundance of these primordial black holes in terms of the bubble nucleation rate at $T_{\rm max}$,  and demonstrate that the abundance can be significantly large from a phenomenological perspective.

\end{abstract}

\maketitle


\section{Introduction}

Primordial black holes (PBHs) are black holes that form in the early Universe in a non-stellar way (see Ref.\,\cite{Escriva:2022duf} for a recent review). 
Their possible existence throughout cosmic history has rich phenomenological implications\,\cite{Carr:2009jm, Carr:2020gox, Carr:2020xqk, Escriva:2022duf, Auffinger:2022khh} 
and a broad mass range of PBHs are compelling candidates for the dark-matter component of the Universe\,\cite{Chapline:1975ojl, Ivanov:1994pa, Carr:2016drx} that might be on the verge of being probed using solar ephemerides precision measurements\,\cite{Tran:2023jci, Loeb:2024tcc}.
Moreover, PBHs could also explain a variety of conundrums, including the recently observed microlensing signal candidates, the correlations in the cosmic infrared and X-ray backgrounds, and the origin of the supermassive black holes in galactic nuclei at high redshift\,\cite{Carr:2019kxo}.
Moreover, it is possible that the LIGO/Virgo black hole mergers\,\cite{LIGOScientific:2016dsl,KAGRA:2021vkt} has a primordial origin\,\cite{Bird:2016dcv}.

So far, most of the PBH formation mechanisms involve the gravitational collapse of large curvature perturbations generated during inflation. To generate such large curvature perturbations, the inflation model is required to have peculiar features, e.g., an inflection point or a plateau in a small field range of the potential\,\cite{Starobinsky:1992ts, Ivanov:1994pa, Ivanov:1997ia, Garcia-Bellido:2017mdw, Ezquiaga:2017fvi, Motohashi:2017kbs, Ballesteros:2017fsr, Cicoli:2018asa, Byrnes:2018txb, Cheong:2019vzl, Ballesteros:2020qam}, a potential hill\,\cite{Yokoyama:1998pt, Briaud:2023eae, Heurtier:2022rhf,Mishra:2019pzq}, multiple phases of inflation or hybrid inflation\,\cite{Garcia-Bellido:1996mdl, Kawasaki:1997ju,Kawasaki:1998vx, Kawasaki:2006zv, Kawaguchi:2007fz, Frampton:2010sw, Kawasaki:2012kn, Clesse:2015wea, Kawasaki:2016pql, Inomata:2016rbd, Inomata:2017okj, Tada:2019amh}, a non-canonical kinetic term\,\cite{Ballesteros:2018wlw, Ballesteros:2021fsp}, multifield inflation\,\cite{Palma:2020ejf, Fumagalli:2020adf}, light spectator fields\,\cite{Yokoyama:1995ex, Kawasaki:2012wr, Kohri:2012yw, Pi:2021dft}, and other possibilities (e.g.\,\cite{Linde:2012bt, Cai:2018tuh,Cai:2019bmk,Heydari:2021gea,Heydari:2021qsr}). 
In addition, PBH formation has been considered in connection with preheating after inflation\,\cite{Green:2000he, Bassett:2000ha, Martin:2019nuw,Martin:2020fgl} although their formation in this context was recently questioned\,\cite{Ballesteros:2024hhq}.

Long after the idea was suggested in Refs.\,\cite{Kodama:1982sf, Hawking:1982ga}, recent works reconsidered that PBHs may also be formed from a first-order phase transition (FOPT)\,\cite{Garriga:2015fdk, Liu:2021svg, Gross:2021qgx, Baker:2021nyl, Kawana:2021tde, Jung:2021mku}. This idea was then further investigated in Refs.\,\cite{Deng:2017uwc, Baker:2021sno,Hashino:2021qoq,  Huang:2022him, He:2022amv, Kawana:2022olo, Lewicki:2023ioy, Gouttenoire:2023naa,Huang:2023chx,Baldes:2023rqv,Gouttenoire:2023pxh,Huang:2023mwy,Lewicki:2024ghw, Flores:2024lng, Kanemura:2024pae, Cai:2024nln, Goncalves:2024vkj,Banerjee:2024cwv}. This possibility is particularly exciting, as FOPTs are naturally present in many particle physics models and have far-reaching phenomenological consequences, such as generating a stochastic gravitational wave background.

In this article, we propose a new PBH formation mechanism in which an FOPT occurs while the Universe's temperature increases during reheating after inflation. This FOPT is thus a heating phase transition\,\cite{Buen-Abad:2023hex, Azatov:2024auq, Barni:2024lkj} rather than a cooling phase transition that occurs as the temperature decreases in the early Universe. The special ingredient of our scenario is an abortion of the FOPT, assuming that the maximal temperature reached in reheating is higher than the critical temperature but lower than the temperature that guarantees the phase transition to complete. In the following, we introduce the specifics of the interrupted phase transition, explain how PBH can form in this setup, and relate the PBH mass and abundance to the dynamics of the perturbative reheating and the phase transition sector considered.

\section{Reheating sector}

Before going into the details, let us be clear about our setup.
When inflation ends, we consider the Universe to be filled with a pressureless fluid slowly decaying into particles that quickly get thermalized, producing a relativistic plasma. 
We refer to this decaying matter component as the {\it reheaton}, $\chi$. 
As $\chi$ decays, the radiation sector's temperature first increases, reaching the maximal temperature $T_{\rm max}$, and decreases as the Universe expands.
The temperature evolution in terms of the scale factor $a$ can be described by\,\cite{Chung:1998rq}
\begin{align}
    T(a)&= c_1 \, T_{\rm max}\left[\left(c_2 \frac{a}{a_{\rm max}} \right)^{-\frac{3}{2}}-\left(c_2 \frac{a}{a_{\rm max}}  \right)^{-4}\right]^{\frac{1}{4}} \,,
    \label{eq:T}
\end{align}
where $a_{\rm max}$ is the scale factor at $T=T_{\rm max}$,  $c_1=2^{6/5}(27^{1/5}5)^{-1/4}\approx 1.30$ and $c_2=2^{6/5}3^{-2/5}\approx 1.48$.

\begin{figure}
    \centering
\includegraphics[width=1\linewidth]{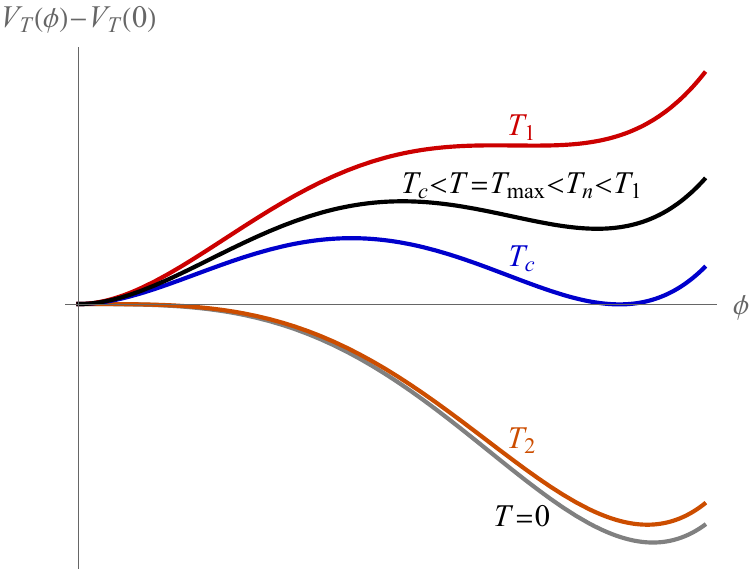}
    \caption{An example of the temperature dependence of scalar potential. }
    \label{fig:HighT-pot}
\end{figure}

Denoting by $\Gamma_\chi$ the decay width of the reheaton, matter domination lasts until the plasma temperature reaches the reheating temperature $T_{\rm RH}\sim \sqrt{\Gamma_\chi M_{\rm Pl}}$, with $M_{\rm Pl}=2.4\times 10^{18}\,\GeV$ being the reduced Planck mass, below which radiation domination starts. 
In general, there is no direct relation between $T_{\rm RH}$ and $T_{\rm max}$, as the value of the latter depends on the time at which the reheating starts.

\noindent
\section{Interrupted phase transition}

Now, let us consider a real scalar field $\phi$ which breaks a symmetry spontaneously, by getting a nonzero vacuum expectation value. 
Assuming that the scalar sector undergoes an FOPT along the temperature change, one can define three characteristic temperatures that play an important role:
the critical temperature, $T_c$ at which two local minima are degenerate, the spinodal (binodal) temperature $T_1$ ($T_2$) above (below) which the potential barrier disappears (see Fig.\,\ref{fig:HighT-pot} for the schematic description of thermal effective potential $V_T(\phi)$ at each temperature). 
During inflation, the temperature is zero, and $\phi$ is stabilized in the symmetry-breaking vacuum, assuming that the inflation scale is not too large compared to the curvature scale of the potential.

While the thermal bath is heated, 
the scalar potential $V(\phi)$ receives thermal corrections and there can be two types of phase transitions in general.
During the change of $T=0\to T_{\rm max}>T_c$, the symmetry-restoring vacuum becomes more stable compared to the symmetry-breaking vacuum, and the phase transition occurs.
This phase transition is called {\it symmetry-restoring}, or heating phase transition (see, e.g. Refs.\,\cite{Buen-Abad:2023hex, Azatov:2024auq, Barni:2024lkj}, for related discussions in various contexts). In previous studies, it is assumed that the heating phase transition is completed and that the Universe settles down in the symmetry-restoring phase. Then, as the temperature drops back, the symmetry-breaking vacuum becomes more stable again, and the symmetry-breaking (or cooling) phase transition starts at the bubble nucleation temperature.

On the contrary, in this paper, we assume that $T_{\rm max}$ is greater than the critical temperature $T_c$, but not large enough to make the bubble nucleation rate catch up with the spacetime expansion. This imposes the condition $T_{\rm max} < T_n \lesssim T_1$, where $T_n$ is the would-be phase transition nucleation temperature if the temperature kept increasing.
Thus, the phase transition is {\it interrupted} at $T_{\rm max}$ by the temperature's turning around.
Bubbles can still be formed, but since they never collide with each other, they just expand during $T>T_c$ and shrink back when $T<T_c$.
We argue that these bubbles eventually lead to PBH formation and that the abundance of such PBHs can be significant.

\section{Fate of bubbles in the interrupted phase transition}

Initially, the bubble grows since the free energy density difference, $\Delta V_T 
\equiv V_T(\phi_b)-V_T(\phi_s)$, is positive, where  $\phi_b$ and $\phi_s$ denote the symmetry-breaking and restoring extrema of the thermal effective potential, respectively.
Once the wall starts expanding, the perturbed plasma would backreact to the wall, creating a backreaction force $\P_{\rm back}(v_w)$, which has a dependence on the wall velocity. 
In general, a terminal velocity exists and should be reached after a short acceleration period, determined by $\Delta V_T=\P_{\rm back}(v_w)$\,\cite{Azatov:2024auq, Barni:2024lkj}.

As the temperature changes, $\Delta V_T$ also changes, so the wall velocity adiabatically follows the terminal velocity at each temperature; it reaches a maximal value at $T_{\rm max}$ and decreases as $T$ decreases. 
At $T=T_c$, $\Delta V_T=0$ which leads to vanishing wall velocity. This is the moment when the bubble stops expanding and has the largest comoving radius, which we denote as $r_{c,2}$.
The subscript ${c,2}$ will be used to indicate quantities estimated at the critical temperature reached for the second time throughout this paper. The critical temperature was reached for the first time during the temperature-increasing process, $T=0 \to T_{\rm max}$, for which we use the labelling of $c,1$.

We can estimate $r_{c,2}$ as $r_{c,2}=\int_{t_{\rm nuc}}^{t_{c,2}} \d t' v_w(t')/a(t')\sim \bar{v}(\eta_{c,2}-\eta_{\rm nuc})$ where $\eta$ is the conformal time defined via $\d t= a\d \eta$, $\bar{v}$ denotes the averaged wall velocity in bubble expansion, and the subscript ${\rm nuc}$ indicates quantities estimated at the time when this bubble is nucleated.
In a matter-dominated universe, we have $H(a)\propto a^{-3/2}$
and thus 
\begin{align}
    r_{c,2} \sim \bar v  (\eta_{c,2} -\eta_{\rm nuc})&= \! \frac{2 \,\bar v}{a_{c,2} H_{c,2}} \!\!
    \left[ \!
    1 \! - \! \left(\frac{a_{\rm nuc}}{a_{c,2}}\right)^{\frac{1}{2}} \! \right].
    \label{eq:rc2}
\end{align}
This shows that the comoving radius at $t_{c,2}$ is of the order of the comoving Hubble radius $r_{H}=1/(a H)$.

Afterwards, at $T<T_c$, the net pressure $\Delta V_T$ becomes negative, and the bubble starts shrinking. The bubble wall velocity stays following its terminal (negative) velocity, which induces fluid motion of the plasma in this region. 
This shrinking occurs slightly below $T_c$ when the vacuum energy difference is comparable to the pressure of the radiation plasma. Therefore, the bubble wall does not run away, as shown in more detail in Appendix~\ref{sup:wall}.
In the absence of any runaway during both its expansion and contraction phases, the energy budget of the bubble wall's kinetic motion is negligible. To understand what happens in the region perturbed by the bubble wall, we can thus focus on the balance between vacuum and thermal energy, where the latter should be understood to include the fluid's bulk motion. During bubble expansion, i.e. when $T>T_c$, the thermal energy is first transferred into vacuum energy, which redshifts more slowly than the radiation plasma with cosmic expansion.
Later on, the bubble's contraction at $T<T_c$ converts vacuum energy back to thermal energy.
Therefore, the energy density of the region perturbed by the wall's motion is greater than the unperturbed region far away from the nucleation site.

Since this perturbation originates from the vacuum energy density difference $|\Delta V_0|$, the initial density contrast $\delta_{\rm i}$ should be roughly given as $\delta_{\rm i} \sim |\Delta V_0|/\bar \rho_\chi(a_{\rm max})$ where $\bar{\rho}_\chi$ is the unperturbed reheaton energy density.
More precisely, there is an additional dependence on the time scale between the bubble's nucleation and disappearance, which is governed by the ratio $a_{c,2}/a_{\rm max}$.
Thus, $\delta_{\rm i}$ can be expressed as
(see Appendix~\ref{sup:criterion} for its derivation)
\bal
\delta_{\rm i} &=
\kappa\Big(\sqrt{\frac{a_{c,2}}{a_{\rm max}}}\Big) \, \frac{|\Delta V_0|}{\rho_\chi (a_{\rm max})}
\label{eq:delta_i}
\eal
where $\kappa(y) = 4 y (y-1)
[(2y-1)^4+1]
[ 1+ (2y-1)^{-2} ]$.
The ratio $a_{c,2}/a_{\rm max}$  depends on the details of the particle physics model that realizes the corresponding phase transition (we provide an example toy model in Appendix~\ref{sup:model} where $a_{c,2}/a_{\rm max} \simeq$ $2$ -- $4$ is realized). 
Note that the ratio $|\Delta V_0|/\bar{\rho}_\chi(a_{\rm max})$ gets suppressed as we take a large $a_{\rm RH}/a_{\rm max}$ because $|\Delta V_0|$ cannot be greater than $\frac{1}{3}\bar{\rho}_{\rm rad}$ at the critical temperature, and the ratio $\bar{\rho}_{\rm rad}(a_{\rm max})/\bar{\rho}_\chi(a_{\rm max}) \simeq (a_{\rm max}/a_{\rm RH})^{3/2}$.
Although small, this overdensity can act as a seed of the PBH formation via the post-collapse accretion mechanism\,\cite{deJong:2021bbo, DeLuca:2021pls,deJong:2023gsx}  (see also Refs.\,\cite{Khlopov:1980mg, Khlopov:1985fch, 
Harada:2013epa, Harada:2016mhb, Harada:2017fjm, Kokubu:2018fxy, Harada:2022xjp, 
Hidalgo:2017dfp, Carr:2017edp, Carr:2018nkm, Allahverdi:2020bys, Martin:2020fgl, Carrion:2021yeh, deJong:2023gsx} for PBH formation during matter domination in a variety of different aspects).

A numerical relativity simulation of the post-collapse accretion mechanism was conducted in Ref.\,\cite{deJong:2021bbo}.
Although the initial setup is different\footnote{
In Ref.\,\cite{deJong:2021bbo}, the initial density perturbation is given by a spherical wave of a massless scalar field and is set to be much larger than in our scenario. 
}, their findings on how the system evolves with a small initial density perturbation in matter domination are still applicable to our case, so we could detail our mechanism as follows.

In the post-collapse accretion mechanism\,\cite{deJong:2021bbo} 
an overdense region of
a macroscopic size 
comparable to the Hubble radius creates a gravitational potential and triggers an accretion of reheaton into this region.
Initially, this accretion leads to linear growth of the density contrast, $\delta(t)\sim \delta_{\rm i} a(t)/a(t_{\rm i})$. As soon as $\delta(t)$ reaches $\delta_{\rm NL}\sim \O(0.1)$, the density contrast grows non-linearly, at an extremely high rate $\delta(t)\sim \delta_{\rm NL} (a(t)/a(t_{\rm NL}))^{34}$\,\cite{deJong:2021bbo}.
This non-linear growth quickly leads to the whole region collapsing into a black hole with an initial mass of order $10^{-2} M_H$\,\cite{deJong:2021bbo} where $M_H=4\pi M_{\rm Pl}^2/H$ is the Hubble mass for a given background expansion rate $H$. 
As shown in Ref.\,\cite{deJong:2021bbo}, after it forms, the black hole quickly increases in mass by absorbing the surrounding matter.
Once the PBH mass reaches about one Hubble mass, the rapid accretion is expected to be slowed down, and the mass simply follows the scaling of one Hubble mass $M_{\rm BH} \sim M_H \propto a^{3/2}$. This mass-growing process ends when radiation domination starts. Eventually, the final PBH mass is simply determined by the value of the Hubble mass at the time of the reheating, which we evaluate by considering that there is a matter-radiation equality at $T_{\rm RH}$, giving
\bal
M_{\rm PBH} \sim 3.5\times 10^{-12}\,M_{\odot}\, \alpha
\Big( \frac{10^5\,\GeV}{T_{\rm RH}} \Big)^{\! 2}
\Big( \frac{100}{g_\star(T_{\rm RH})} \Big)^{\! 1/2} \!\! ,
\label{eq:MPBH}
\eal
where $g_\star (T_{\rm RH})$ is the number of effective relativistic degrees of freedom present in the plasma at reheating, and $\alpha \lesssim 1$ is an efficiency factor, which we take to be {$\mathcal O(0.1)$} for simplicity. 
As can be seen from Eq.\,\eqref{eq:MPBH}, $M_{\rm PBH}$ is insensitive to the phase transition properties and solely determined by the value of $T_{\rm RH}$ once they are formed. The distribution of PBHs formed from an interrupted phase transition is thus expected to be monochromatic.

\begin{figure*}
    \centering
    \includegraphics[width=1\linewidth]{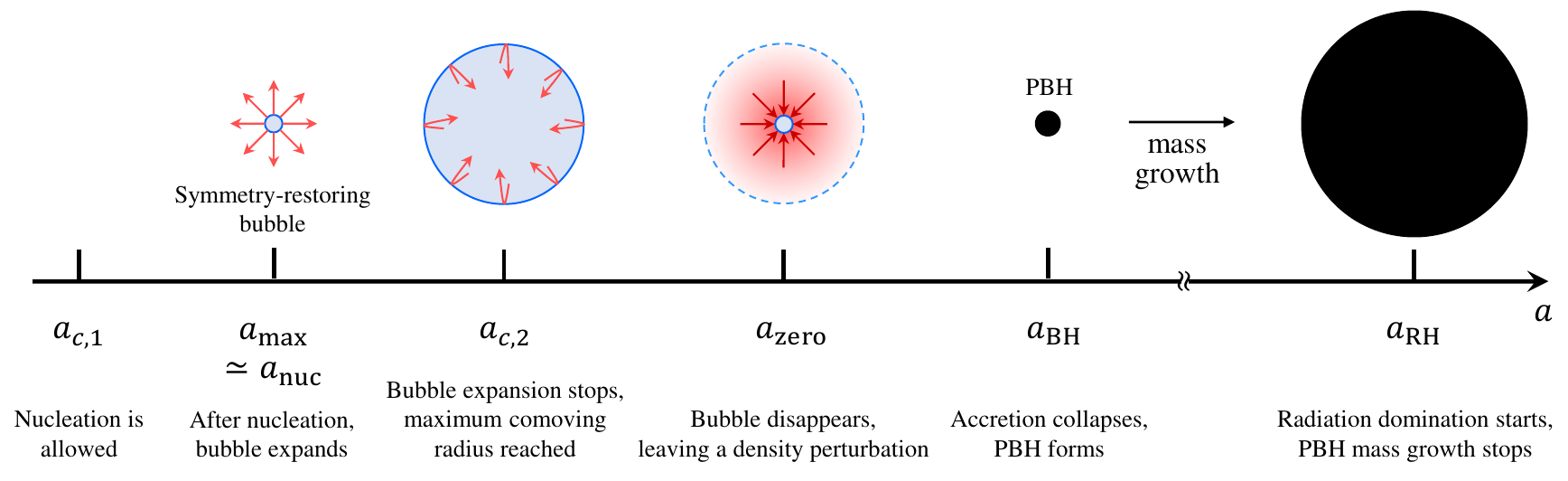}
    \caption{
    A schematic chronology of our PBH formation scenario. A symmetry-restoring bubble nucleates at around $a_{\rm max}$ and expands with the bubble wall indicated by the blue line. 
    At $a_{c,2}$, the bubble wall stops expanding, turns around, and shrinks until it completely disappears at $a_{\rm zero}$.
    This leaves a spherical overdense region of macroscopic size (dashed blue line). 
    This region accretes surrounding matter (reheaton), and the accretion collapses into a PBH via the post-collapse accretion mechanism at $a_{\rm BH}$.
    The PBH mass grows until the radiation domination starts at $a_{\rm RH}$.
    }
    \label{fig:chronology}
\end{figure*}

For the post-collapse accretion mechanism to work for our scenario, the period of linear growth must end before $t_{\rm RH}$.
The scale factor $a_{\rm NL}$ when the non-linear growth begins can be estimated by $\delta_{\rm i} \times (a_{\rm NL}/a_{\rm max}) \simeq \delta_{\rm NL}$ with Eq.\,\eqref{eq:delta_i}.
This sets the following constraint between the particle physics model and reheating temperature $T_{\rm RH}$
\begin{equation}
    \left(\frac{a_{\rm RH}}{a_{\rm max}}\right) <\left(\frac{10 g_{\star,\phi}(T_c)}{3 g_\star(T_{\rm RH})}\right)^{\! 2}  
    \, \, 
    h  \!\left(\frac{a_{c,2}}{a_{\rm max}}\right) \,,
\end{equation}
where $g_{\star,\phi}$ is the number of effective degrees of freedom that strongly couple to $\phi$ resulting in $|\Delta V_0|\simeq \frac{\pi^2}{90} g_{\star,\phi}(T_c) T_c^4$, and the model-dependent function $h(x)=x(2-1/\sqrt{x})^8$ is $O(10)$ -- $O(10^2)$ for $x=2$ -- $4$.
As detailed in Appendix~\ref{sup:criterion}, this constraint is satisfied for a large range of the parameters, including the benchmark parameters used later on in this work, ensuring the validity of our scenario.

As a brief summary of our scenario, we show a schematic picture in Fig.\,\ref{fig:chronology}. Above the first critical temperature, a symmetry-restoring bubble is nucleated around the scale factor $a_{\rm max}$. The bubble expands until the temperature reaches the critical temperature again, shrinks back afterwards, and disappears at $a_{\rm zero}$, leaving a spherical overdense region of macroscopic size (dashed blue line). This region accretes surrounding matter (reheaton), and the accretion collapses into a PBH at $a_{\rm BH}$. The PBH mass continuously grows by absorbing surrounding matter until the radiation domination starts at $a_{\rm RH}$.

\section{PBH abundance}

The PBH relic abundance can be estimated by counting the expected number of symmetry-restoring bubble nucleations during the interrupted phase transition.
It is thus sensitive to the bubble nucleation rate per unit volume, $\Gamma(T)\sim T^4 \e^{-S_3/T}$ where $S_3$ is the minimal energy of the scalar configuration to make a thermal escape from the local minimum, which can be obtained by the three-dimensional Euclidean action of the $O(3)$ bounce solution\,\cite{Linde:1980tt,Linde:1981zj,Gould:2021ccf}. Since the phase transition is aborted, $\Gamma$ is maximized at the moment where $T=T_{\rm max}$, and most of the symmetry-restoring bubbles are nucleated around this time.

To be specific, let us consider a sufficiently large comoving total volume $\overbar{V}$. 
The number of nucleated bubbles at time $t_{\rm nuc}$, corresponding to $a_{\rm nuc}$, is given by 
\begin{align}
    \d N_{\rm PBH} (a_{\rm nuc})= \frac{\d a_{\rm nuc}}{a_{\rm nuc} H(a_{\rm nuc})} \times \overbar{V} a_{\rm nuc}^3 \Gamma(T(a_{\rm nuc}))\,.
\end{align}
Integrating it from $t_{c,1}$ to $t$ and dividing the result by $\overbar{V}a(t)^3$ gives the integrated number density at $t$ 
\begin{align}
\label{eq:nPBH}
    n_{\rm PBH}(t)=\left(\frac{a_{\rm max}}{a(t)}\right)^3  \int_{a_{c,1}}^{a(t)} \left(\frac{a_{\rm nuc}}{a_{\rm max}}\right)^2 \frac{\Gamma(T(a_{\rm nuc}))\d a_{\rm nuc}}{a_{\rm max} H(a_{\rm nuc})}\,.
\end{align}
From this, one can obtain the PBH dark matter fraction $f_{\rm PBH}=n_{\rm PBH}(t_{\rm today}) M_{\rm PBH}/(\rho_c\,\Omega_{\rm DM})$. Here, we proceed with a rough estimation using a model-independent approach with the following approximations.
First of all, we take the Taylor expansion of $S_3/T$ around $T_{\rm max}$ in the $\log$ scale;
\bal
\frac{S_3}{T} \simeq 
\left. \frac{S_3}{T}\right|_{T=T_{\rm max}}
\!\!\! 
- \hat \beta_{\rm max} \ln \left(\frac{T}{T_{\rm max}}\right)\,,
\eal
with $\hat \beta_{\rm max}  \equiv - \left. \frac{\d (S_3/T)}{\d\ln T}\right|_{T=T_{\rm max}}\,.$
Then we can approximate $\Gamma(T) \simeq \Gamma(T_{\rm max}) (T/T_{\rm max})^{\hat \beta_{\rm max}+4}$.
In Appendix~\ref{sup:model}, we evaluate $S_3/T$ and $\hat \beta_{\rm max}$ in the case of the so-called Abelian Higgs model and obtain $\hat \beta_{\rm max}$ around $10^4$--$10^6$. We use this value as a benchmark in what follows. 
In addition, using Eq.\,\eqref{eq:T} to evaluate $T(a)$, we obtain in the limit $a\approx a_{\rm max}$
\begin{align}
    T(a)
    \simeq
    T_{\rm max} \exp \Big[-\frac{3}{4} \Big( \frac{a}{a_{\rm max}}-1 \Big)^2 \Big]\,.
    \label{eq:T_approx}
\end{align}
{Although \eqref{eq:T_approx} and \eqref{eq:T_approx} are only valid around $T_{\rm max}$, we checked numerically that they lead to a good approximation for  $f_{\rm PBH}$ as long as $\hat \beta_{\rm max}>50$ because the largest contribution to $f_{\rm PBH}$ comes from $\Gamma(T_{\rm max})$.} 

\begin{figure}[t]
    \centering
    \includegraphics[width=\linewidth]{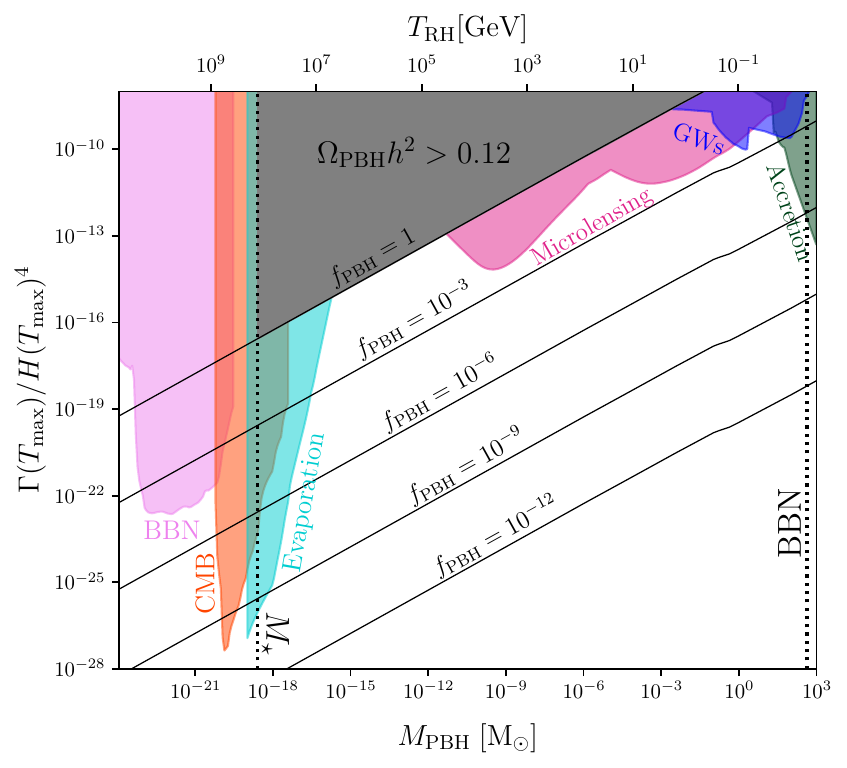}
    \caption{\label{fig:pheno} Fraction of the dark matter relic density that is composed of PBHs, as a function of the PBH mass or equivalently the reheating temperature, using the benchmark values {$\alpha=0.1$}, {$a_{\rm RH}/a_{\rm max}=10$} and $\hat\beta_{\rm max}=10^5$. Shaded areas correspond to regions of the parameter space excluded by BBN, CMB anisotropies, cosmic-ray detection, microlensing, gravitational wave detection and accretion (as reported in~\cite{Carr:2020gox, Carr:2009jm}).
    }
     
\end{figure}

Then, Eq.\,\eqref{eq:nPBH} can be approximated as
\bal
n_{\rm PBH}(t_{\rm RH})
\simeq
\sqrt{ \frac{4\pi}{3\hat{\beta}_{\rm max}}} 
\bigg(\frac{a_{\rm max}}{a_{\rm RH}}\bigg)^3
\frac{\Gamma(T_{\rm max})}{H_{\rm max}},
\eal
for $\hat \beta_{\rm max}>50$.
Assuming that the PBH yield is unchanged after reheating temperature, we obtain $f_{\rm PBH}$ as
\bal
&f_{\rm PBH}  = \frac{M_{\rm PBH} n_{\rm PBH}/s}{\rho_{\rm DM}/s}
\notag
\\
&\sim
1\,\alpha \,
\bigg( \! \frac{T_{\rm RH}}{10^5\,\GeV} \! \bigg)
\Bigg( \!  \frac{ \left( \frac{\Gamma(T_{\rm max})}{H_{\rm max}^4} \! \right)}{10^{-17}} \Bigg)
\bigg( \frac{10^4}{\hat\beta_{\rm max}} \bigg)^{\!\! \frac12}
\!
\bigg( \frac{a_{\rm RH}/a_{\rm max}} {10}\bigg)^{\!\! \frac32} \! ,
\label{eq:fPBH}
\eal
where the observed dark matter relic abundance is taken to be $\rho_{\rm DM}/s \simeq 0.4\,\eV$\,\cite{Planck:2018vyg}.

In Fig.\,\ref{fig:pheno}, we depict $\Gamma(T_{\rm max})/H_{\rm max}^4$ required to give a sizable $f_{\rm PBH}$ for different $M_{\rm PBH}$ (or $T_{\rm RH}$) for $\alpha=0.1$, {$a_{\rm RH}/a_{\rm max}=10$} and $\hat{\beta}_{\rm max}=10^5$, taking $g_*(T)$ to be the Standard Model value\,\cite{Saikawa:2018rcs}.
We also show relevant constraints coming from the null observation of PBH evaporation signal (cyan), lensing by PBHs (purple), gravitational waves (blue), and accretion (green), taken from Ref.\,\cite{Auffinger:2022khh}. 
The dotted line on the right edge represents the lower bound of $T_{\rm RH} \gtrsim 5\,\MeV$ (and thus an upper bound of $M_{\rm PBH}$) coming from the big bang nucleosynthesis\,\cite{Kawasaki:1999na, Kawasaki:2000en, Hannestad:2004px, Ichikawa:2005vw,deSalas:2015glj, Hasegawa:2019jsa} while the one on the left edge depicts the critical PBH mass $M_\star\simeq 5\times 10^{14}\,{\rm gram}$ below which PBHs evaporate completely before the present\,\cite{Carr:2009jm, Carr:2016hva}.
For masses smaller than $M_\star$, we also indicate constraints from BBN (pink) and CMB anisotropies (orange)\,\cite{Auffinger:2022khh}. 
Remarkably, a broad range of values for $\Gamma(T_{\rm max})/H_{\rm max}^4$ lead to an abundance of PBHs that is of phenomenological interest, including PBHs that could constitute the whole dark matter of our Universe.

\vspace{0.2cm}

\section{Discussion}

In this work, we have proposed a new PBH formation mechanism in an interrupted phase transition during reheating.
A symmetry-restoring bubble is nucleated and expands during $T>T_c$, and it shrinks back as the temperature drops below $T_c$.
This generates a macroscopic size of over-density perturbation with a spherical symmetry, which eventually collapses into a PBH via the post-collapse accretion mechanism during matter domination.
The mass of PBHs formed in this process grows quickly by absorbing the surrounding matter, and its final mass is determined by $T_{\rm RH}$ as given in Eq.\,\eqref{eq:MPBH}.
We estimate the PBH abundance \eqref{eq:fPBH} in terms of the bubble nucleation rate around $T_{\rm max}$ parametrized by the effective rapidity parameter $\hat \beta_{\rm max}$ at $T_{\rm max}$, and show that it can be sizable in the aspect of phenomenology.

Our findings rely on the {post-collapse accretion} mechanism\,\cite{deJong:2021bbo, DeLuca:2021pls, deJong:2023gsx}, in which a small overdensity accretes the matter present in the Hubble patch during a matter-dominated era, leading to the formation of a black hole. 
The estimation of PBH number density assumes a high sphericity of the initial density perturbation\,\cite{Harada:2016mhb}, and that the surrounding matter (reheaton) has small inhomogeneity\,\cite{Kokubu:2018fxy} and velocity dispersion\,\cite{Harada:2022xjp}.
In the following remarks, we discuss qualitative justification of these assumptions and potential issues that need further investigation.

Since the initial density perturbation originates from a symmetry-restoring bubble in our scenario, let us explain why we expect bubbles to have high sphericity.
Initial bubbles at the nucleation do have a small non-sphericity as a thermal noise about the $O(3)$-symmetric critical bubble profile, since our bubbles are nucleated via thermal escapes.
This initial non-sphericity at the nucleation gets suppressed and become negligible during the bubble expansion and contraction because (i) the bubble wall has a unique terminal velocity determined by the balance of the vacuum pressure and the particle-wall interaction, (ii) the surface tension on the bubble wall stabilizes the bubble shape at a spherical configuration, and (iii) any scalar field vibration on the wall would quickly be dissipated into the plasma due to the sufficiently large friction. 
In some limited cases for subsonic bubble wall\,\cite{Link:1992dm, Megevand:2013yua}, an instability of the spherical deflagration may arise.
In our case, bubbles are supersonic most time, so they are safe from this instability.
The turnaround moment may be the only chance of this instability, while the effect cannot be large because the bubble at that time is already macroscopic, and microscopic fluctuation cannot deform it macroscopically.
However, we note that further investigation with quantitative estimation may be needed.

The matter to be accreted into the region perturbed by a bubble, the reheaton, may possess small but nonzero inhomogeneities and velocity dispersion inherited from inflation, which may impede the formation of the black hole, and its mass growth partially.
These inhomogeneities, and the associated departures from sphericity, may be amplified during the post-collapse accretion phase. Such effects are generic to primordial black hole formation in a matter-dominated era and are not specific to the mechanism considered here. A detailed analysis is therefore beyond the scope of this paper, and we defer it to future work.

\begin{acknowledgments}
We thank Shao-Jiang Wang for the helpful discussions. The work of WYA was supported by  EPSRC [Grant No. EP/V002821/1]. The work of LH is supported by the STFC (grant No. ST/X000753/1).
The work of THJ was supported by IBS under the project code, IBS-R018-D1.
\end{acknowledgments}


\appendix




\section{An example model to evaluate the phase transition rapidity parameter}
\label{sup:model}
In this section, we consider a benchmark model and obtain $\hat{\beta}_{\rm max}$.
The model we consider is a simple Abelian Higgs model where a complex scalar field $\Phi$ is charged under a $U(1)$ gauge interaction with a charge unity.
We further assume that the theory is classically scale invariant, so the tree-level potential is given by
\bal
V_{\rm tree}(\Phi) = \lambda |\Phi|^4 +{\rm const},
\eal
where $\lambda$ is the self-quartic coupling and the constant is introduced to ensure that the potential is zero at the broken phase at zero temperature.
The spontaneous symmetry breaking is radiatively generated as originally shown in Ref.\,\cite{Coleman:1973jx}. 

To include the loop effects conveniently, we take the RG scale $\mu=\mu_*$ which is defined by $\lambda(\mu_*)=0$.
The existence of such $\mu_*$ is guaranteed by the positive beta function of $\lambda$ coming from the gauge boson loop.
Denoting $\phi$ for the radial degree of $\Phi$, the one-loop effective potential {at zero temperature} can be written as
\bal
{V_0(\phi)} = \frac{\delta \lambda}{4}\phi^4 + \frac{1}{4}\beta_{\lambda} \phi^4 \log \frac{\phi}{\mu_*},
\eal
where $\delta \lambda = \frac{3\,g^4}{16\pi^2}(\log g^2 - 5/6)$ and $\beta_{\lambda}=6g^4/16\pi^2$ with $g$ being the gauge coupling.
This potential is minimized at $v_\phi = \e^{1/6}\mu_*/g$ and the {zero-temperature} potential energy difference {between the broken phase and symmetric phase} is given by $\Delta V_0 \equiv V_0(\phi_b)- V_0(\phi_s)  =  - \frac{3\,\e^{2/3}}{128\pi^2} \mu_*^4$.

We include the thermal correction coming from the gauge boson loop, 
\bal
\delta V_{T\neq 0}  = \frac{3T^4}{2\pi^2}
J_B(m_V^2/T^2),
\eal
with the field-dependent gauge boson mass $m_V=g\phi$ and the $J_B$ function given by
\bal
J_B(y^2)
=
\int_0^\infty
\d x \, x^2 \log\left[
1-\e^{-\sqrt{x^2+y^2}}
\right].
\eal
Note that the scalar-loop contribution vanishes due to our choice of RG scale, $\lambda(\mu_*)=0$. In our notation, $V_T=V_0+\delta V_{T\neq 0}$.

In this specific setup, we find two important model properties.
First, $T_c$ is independent of the size of gauge coupling $g$ since the zero-temperature potential energy difference is given by $|\Delta V_0| = \frac{3\,\e^{2/3}}{128\pi^2} \mu_*^4$ independently of $g$.
Second, the binodal temperature $T_1$ (where the potential barrier disappears) is also $g$-independent.
This is because the effective field range of the thermal correction $\phi_{{\rm eff},\,T}$ has the same coupling dependence with $v_\phi\sim \mu_*/g$; $\phi_{{\rm eff},\,T}$ can be estimated by $m_V(\phi_{{\rm eff},\,T}) \sim T$, so $\phi_{{\rm eff},\,T} \sim T/g$.
We numerically find that $T_c\simeq 0.37 \, \mu_*$
and $T_1 \simeq 0.44 \,\mu_*$.
For our PBH formation scenario, $T_{\rm max}$ must be between $T_1$ and $T_c$, which is not impossible, although it requires tuning (note that there are already multiple coincidences of time scales in the standard cosmology). 

\begin{figure}[t]
    \centering
    \includegraphics[width=0.47\textwidth]{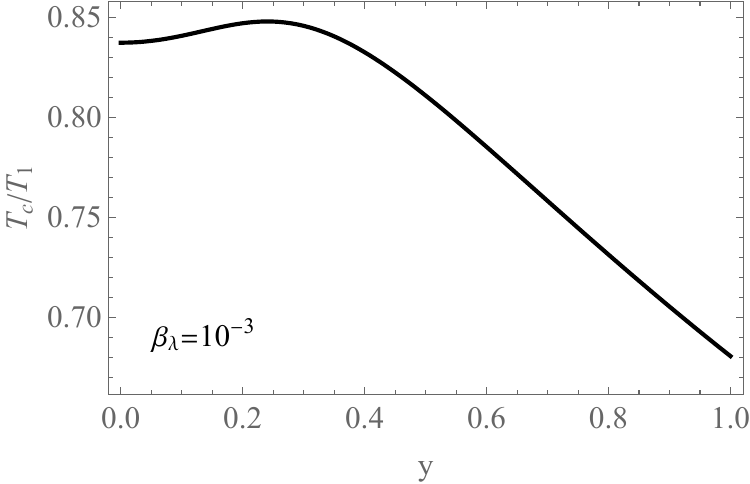}
    \includegraphics[width=0.47\textwidth]{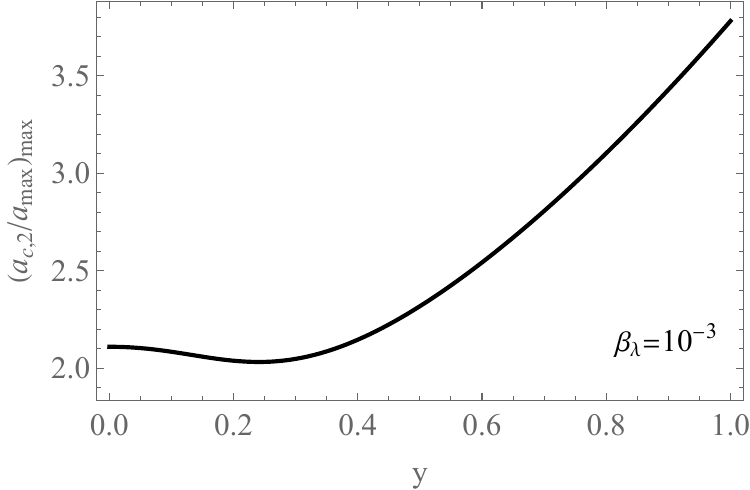}
    \caption{$T_c/T_1$ (left) and corresponding $(a_{\rm max}/a_{c,2})_{\rm min}$ (right) are depicted when fermions are introduced with Yukawa interactions. Gauge coupling is chosen such that $\beta_\lambda =10^{-3}$.}
    \label{fig:T1overTc}
\end{figure}

These properties can be changed by including additional fields. For instance, $T_1/T_c$ can be increased when we include Weyl fermions $\psi_i$ and $\chi_i$ with $i=1,2$ that couple to $\phi$ via Yukawa interactions.
For the gauge anomaly cancellation, we take the gauge charges of $\psi_1$ and $\psi_2$ oppositely $\pm 1$ while $\chi_i$ has no charge under the gauge interaction.
Then, the Yukawa interaction becomes $y_1 \Phi \psi_1^\dagger \chi_1 + y_2 \Phi \chi_2^\dagger \psi_2 + {\rm h.c.}$.
For simplicity, we ignore a potential flavor structure by taking $y_1=y_2=y$ and obtain $T_c$ and $T_1$ numerically. 
Then, $T_c/T_1$ is depicted in the left panel of Fig.\,\ref{fig:T1overTc}, where we fix the gauge coupling such that $\beta_{\lambda}=\frac{1}{16\pi^2}(6g^4-2 y^4) =10^{-3}$.
From $T_c/T_1$, we can estimate the minimal value of $a_{\rm max}/a_{c,2}$ that the model can reach. Because $T_{\rm max}$ must lie between $T_1$ and $T_c$ for the interrupted phase transition, $T_{\rm max} \approx
T_1$ will give the smallest value of $a_{\rm max}/a_{c,2}$. Assuming Eq.\,\eqref{eq:T}, we depict it on the right panel of Fig.\,\ref{fig:T1overTc}.

In this example model, as we increase $y$, the parametric tuning between the gauge coupling and Yukawa coupling contributions inside $\beta_{\lambda}$ gets severe.
For $\beta_{\lambda}=10^{-3}$ and $y=1$, the level of tuning (which can be defined by $\beta_{\lambda} \left( \frac{6g^4}{16\pi^2}\right)^{-1}$) becomes $7\,\%$.
If we take $y$ even larger (or equivalently, smaller $\beta_\lambda$ with fixing $y$), the two-loop contribution becomes more and more important, and eventually, our calculation becomes invalid.
This is because there is no symmetric argument that guarantees the cancellation at a higher loop order. However, of course, the cancellation structure can be provided by a symmetric reason in some models, e.g. supersymmetry.

\begin{figure}[t]
    \centering
    \includegraphics[width=1\linewidth]{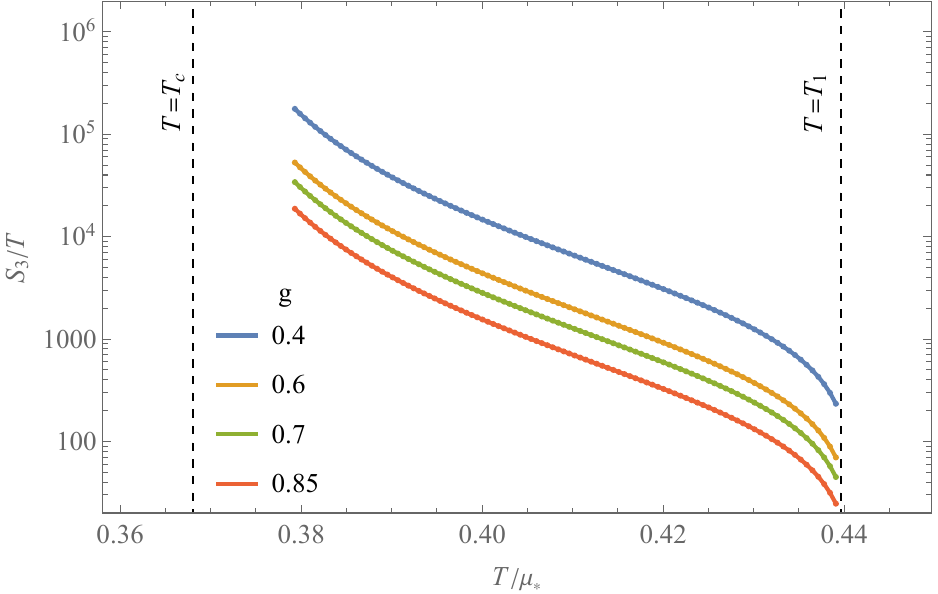}
    \includegraphics[width=1\linewidth]{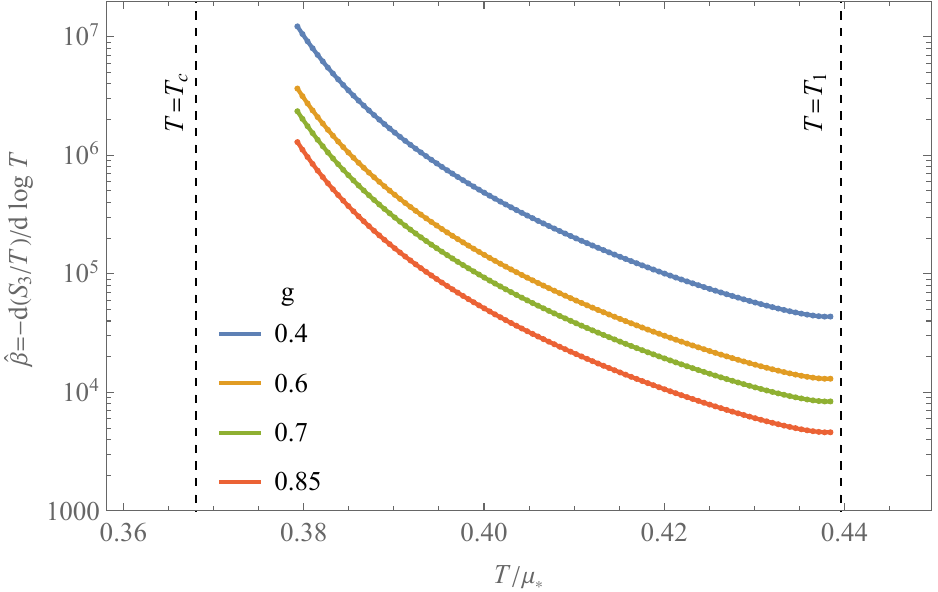}
    \caption{Our numerical result of $S_3/T$ and $\hat \beta=-\d(S_3/T)/\d \ln T$ for the Abelian Higgs model. Here, we take $y=0$.}
    \label{fig:CW_S3overT}
\end{figure}

In fact, in order to avoid bubble collisions, we require 
\begin{align}
\label{eq:constriantTmax}
    \Gamma(T_{\rm max})\sim T_{\rm max}^4\,\e^{-S_3(T_{\rm max})/T_{\rm max}} < H(T_{\rm max})^4\,,
\end{align}
where $\Gamma(T)$ is the bubble nucleation rate and $S_3$ is the bounce action\,\cite{Linde:1980tt,Linde:1981zj}.
This condition means that the average number of bubbles per Hubble volume never reaches one and therefore our scenario of PBH formation involves only single-bubble dynamics. Using $H(T_{\rm max})^4=H(T_{\rm RH})^4 (a_{\rm RH}/a_{\rm max})^6$ and 
\begin{align}
    \left(\frac{a_{\rm RH}}{a_{\rm max}}\right)\approx \frac{4}{5^{2/3}} \left(\frac{T_{\rm max}}{T_{\rm RH}}\right)^{\frac{8}{3}} 
\end{align}
from Eq.~\eqref{eq:T}, we can write 
\begin{align}
\label{eq:H(Tmax)}
    H(T_{\rm max})^4&=H(T_{\rm RH})^4\left(\frac{4}{5^{2/3}}\right)^6 \left(\frac{T_{\rm max}}{T_{\rm RH}}\right)^{16}\notag\\
    &=\left(\frac{2\pi^2}{90 M_{\rm Pl}^2} g_{\star}(T_{\rm RH}) T^4_{\rm RH})\right)^2\left(\frac{4}{5^{2/3}}\right)^6 \left(\frac{T_{\rm max}}{T_{\rm RH}}\right)^{16}\,,
\end{align}
where we have used the matter-radiation equality at $T_{\rm RH}$. Note that the approximation $(a/a_{\rm max})\approx 4/5^{2/3} (T_{\rm max}/T)^{8/3}$ works well already for $a>2 a_{\rm max}$. And for typical models, we have $a_{\rm RH}> a_{c,2}> 2 a_{\rm max}$.
Substituting Eq.\,\eqref{eq:H(Tmax)} into Eq.\,\ref{eq:constriantTmax}, we obtain
\begin{align}
    \frac{S_3(T_{\rm max})}{T_{\rm max}} \gtrsim 121 + 4\log\left(\frac{10^5\,{\rm GeV}}{T_{\rm max}}\right)+ 8\log\left(\frac{T_{\rm RH}}{T_{\rm max}}\right)\,,
\end{align}
where we have assumed $g_\star(T_{\rm RH})\approx 100$. This provides an upper bound for $T_{\rm max}$, which we denote as $T_n$. Apparently, $T_n$ is the nucleation temperature for a heating FOPT in a matter-dominated universe.
For example, if $T_{\rm max}\gtrsim T_{\rm RH}\sim 10^{5} {\rm GeV}$, we then have $S_3(T_{\rm max})/T_{\rm max} \gtrsim 121$. From the left panel of Fig.~\ref{fig:CW_S3overT}, we can read that for the particle physics model under consideration, $T_n\approx T_1$.

For a temperature between $T_c$ and $T_1$, we obtain the bounce action by using the CosmoTransitions\,\cite{Wainwright:2011kj}.
The result of $S_3/T$ is given in the left panel of Fig.\,\ref{fig:CW_S3overT} where we now turn off the Yukawa coupling.
Then, we obtain the rapidity parameter as shown in the right panel of Fig.\,\ref{fig:CW_S3overT}, which shows that $10^4 \lesssim \hat{\beta}_{\rm max} \lesssim 10^6$. 

\section{Dynamics of a symmetry-restoring bubble}
\label{sup:wall}

In this section, we show that the bubble wall typically reaches a terminal velocity, i.e., has a non-runaway behavior, in both the expansion and contraction stages.
Bubble wall dynamics is a highly complicated subject, requiring one to solve the Boltzmann equations for the particle distribution functions (which are integro-differential equations), the background scalar equation of motion, and the fluid equations for the hydrodynamics\,\cite{Liu:1992tn,Moore:1995ua, Moore:1995si,Konstandin:2014zta,Ai:2021kak,Laurent:2022jrs,Ekstedt:2024fyq}. To determine whether or not a bubble wall runs away, i.e. accelerates all the way until colliding with another bubble, friction in the $\gamma_w\rightarrow \infty$ limit, where $\gamma_w$ is the Lorentz factor of the wall velocity, is usually compared to the vacuum energy difference\,\cite{Bodeker:2009qy} (although this may not always be valid\,\cite{Ai:2024shx}).
Below, we also do a similar analysis, using the simple Bödeker-Moore criterion\,\cite{Bodeker:2009qy}.

\subsection{Bubble expansion (\texorpdfstring{$T>T_c$}{TEXT})}

Let us first consider bubble expansion. 
Note that, for $T>T_c$, the vacuum energy inside the bubble is greater than outside, i.e. $\Delta V_0 <0$, so the vacuum energy always gives a negative pressure that tries to contract the bubble.
On the other hand, the thermal pressure difference $\Delta V_{T\neq 0}$ is positive, and this is the driving force of the bubble expansion.
When a bubble is formed at $T>T_c$, the bubble wall gets accelerated since the net pressure is positive (by the definition of $T_c$).
When the bubble wall velocity is nonzero, the thermal driving force is reduced.
This can be seen from the fact that, in the example of 1-to-1 transmission processes, the momentum transfer in the wall-rest frame decreases as the fluid velocity increases; $\Delta p_z = \sqrt{p_z^2 + \Delta m^2} - p_z \sim \Delta m^2/2p_z$ where $z$ is the direction of the bubble wall propagation, $p_z$ is the momentum of a particle coming toward the bubble wall from outside, and $\Delta m^2>0$ is the mass-squared difference.
Therefore, as velocity increases, the thermal driving force decreases until it reaches the equilibrium with the vacuum energy pressure.

As pointed out in Ref.~\cite{Barni:2024lkj}, the thermal driving force has a nonzero asymptotic value in $\gamma_w \to \infty$ limit, which we also call Bödeker-Moore thermal force
\begin{align}
     \P_{\rm BM}=\sum_i C_i g_ic_i\frac{\Delta m_i^2 T^2}{24}\,,
\end{align}
where $c_i=1 (1/2)$ for bosons (fermions). Here $g_i$ is the number of internal degrees of freedom of species $i$ that couple with the scalar $\phi$, and $\Delta m_i^2$ is the difference of the squared-mass in broken and symmetric phases.  $C_i$ is approximately given by
\begin{align}
   \frac{C_i  T^2}{24} \approx  
\begin{cases}
\frac{T^2}{24}\quad &\text{if} \quad m_i^{\rm out} \ll T\;,
\\
\frac{1}{2 m_i^{\rm out }}\bigg(\frac{m_i^{\rm out} T}{2\pi}\bigg)^{3/2} \e^{-m_i^{\rm out}/T}\quad &\text{if} \quad   m_i^{\rm out}\gg T\,,
\end{cases}  
\label{eq:ceff}
\end{align}
with $m_i^{\rm out}$ being the mass outside of the wall, i.e. in the broken phase.

If $m_i^{\rm out}$ are larger than the temperature, which is the case for our model considered in the last section, $\P_{\rm BM}$ would be suppressed because the number density of those heavy particles is Boltzmann-suppressed. 
This means that the asymptotic value of the driving force is small, ensuring the existence of equilibrium with $\Delta V_0$ at some velocity.

We note that $\P_{\rm BM}$ is the force caused only by the $1\rightarrow 1$ processes. There can be additional forces caused by particle-production processes\,\cite{Bodeker:2017cim, Hoche:2020ysm, Azatov:2020ufh, Gouttenoire:2021kjv,Ai:2023suz,Azatov:2023xem,Long:2024sqg}, i.e., when a particle splits into two or more particles when it transits across the wall. These next-to-leading-order forces may behave as true friction as in a cooling phase transition\,\cite{Azatov:2024auq}. We also note that hydrodynamic effects can induce a barrier of the frictional pressure at the Jouguet velocity\,\cite{Cline:2021iff, Laurent:2022jrs, Ai:2023see, Ai:2024shx,Ai:2024btx}. All these factors would just make our conclusion more solid.

\subsection{Bubble contraction (\texorpdfstring{$T<T_c$}{TEXT})}

For $T<T_c$, the dynamics of the bubble wall can be understood in the usual way although our bubble is still symmetry-restoring and contracts. Actually, the contraction process under consideration can be likened to the contraction of a {\it false-vacuum} bubble (sometimes referred to as a false-vacuum island) in a cooling and symmetry-breaking FOPT. 
During contraction, the vacuum energy difference accelerates the bubble wall velocity while the thermal effect acts as friction.
In this case, when the wall velocity increases, the friction increases and has an asymptotic value of ${\cal P}_{\rm BM}$\,\cite{Bodeker:2009qy}.
Thus, if $|\Delta V_0| < {\cal P}_{\rm BM}$, there exists a terminal velocity where the friction and $\Delta V_0$ make an equilibrium.

Before proceeding, note that the temperature range in our process is all around $T_c$.
As shown in the previous section, $T_{\rm max}/T_c < T_1/T_c$ cannot be large in the model-building aspect, and therefore, the temperature when the bubble shrinks and disappears, which we denote $T_{\rm zero}$, should be also close to $T_c$.

Now let us again consider the large-$\gamma_w$ limit. In bubble contraction, the driving force is 
\begin{align}
    \P_{\rm driving}=|\Delta V_0|\,,
\end{align}
while the Bodeker-Moore thermal friction is\,\cite{Bodeker:2009qy}
\begin{align}
    \P_{\rm friction}=\P_{\rm BM}=\sum_i g_ic_i\frac{\Delta m^2 T^2}{24}\sim g_{\star,\phi} \frac{\Delta m^2 T^2}{24}\,,
\end{align}
where $g_{\star,\phi}$ is the effective degrees of freedom that strongly couple to $\phi$.

On the other hand, we have the relation of
$|\Delta V_0| \simeq g_{\star,\phi} \frac{\pi^2}{90}T_c^4$ which is smaller than $\P_{\rm BM}$ for $\Delta m^2 > T^2$.
Therefore, we conclude that the bubble wall still does not run away even without taking into account friction from 1-to-2 or 1-to-many processes and hydrodynamic obstruction\,\cite{Ai:2024shx}.

\section{Overdensity generated by a disappearing bubble and criterion for PBH formation}
\label{sup:criterion}

In this section, we carefully analyze the overdensity generated by bubble expansion and contraction and the criterion of successful PBH formation via the post-collapse accretion mechanism.

\subsection{Initial density contrast \texorpdfstring{$\delta_{\rm i}$}{TEXT} generated by the bubble}
We start by looking at a point $p$ with a (comoving) radial distance $0<r<r_{c,2}$ away from the centre of the perturbed region. The overdensity $\delta(r)$ will depend on $r$ but we will take the $r\rightarrow 0$ result as a characteristic value which should be in the same order as the averaged density contrast in magnitude.

\begin{figure}
    \centering  \includegraphics[width=0.5\linewidth]{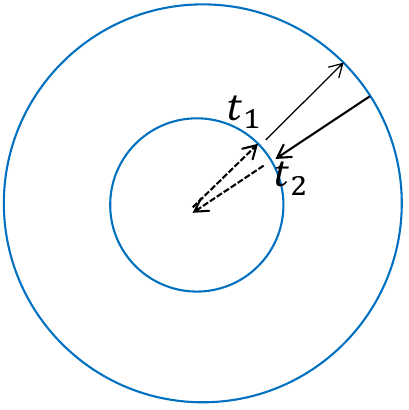}
    \caption{Illustration of how the wall perturbs a position at a distance $r$ from the bubble centre. We compute the density contrast $\delta$ at the centre of the perturbed region by taking $r\rightarrow 0$. }
    \label{fig:bubble-expansion-contraction}
\end{figure}

Once we have a bubble nucleated at the centre at $t_{\rm max}$ (recall that the initial microscopic bubble size is negligible compared to the size of the perturbed region), the bubble wall expands outwards, and will pass the point $p$ at a time denoted by $t_1$. The bubble stops expansion at $t_{c,2}$, reaching its maximal comoving radius $r_{c,2}$, and turns around for contraction. Then the bubble wall will pass $p$ for the second time at a time denoted by $t_2$, See Fig.\,\ref{fig:bubble-expansion-contraction}. For $r\rightarrow 0$, $t_1\rightarrow t_{\rm max}$ and $t_2$ is such that $\eta_2=2\eta_{c,2}-\eta_{\rm max}$.
Actually, it is more convenient to use the cosmological scale factor as the time variable.
Using $H=H_{\rm max}(a_{\rm max}/a)^{3/2}$ for a matter-dominated universe, we have 
\begin{align}
    \left(\frac{a(\eta)}{a_{\rm max}}\right) =  \left[\frac{1}{2}
    {a_{\rm max}}
    H_{\rm max}(\eta-\eta_{\rm max})+1\right]^2 \notag\\ \Rightarrow\quad \eta_{c,2}-\eta_{\rm max}=\frac{2}{
    {a_{\rm max}} 
    H_{\rm max}}
    \left[\left(\frac{a_{c,2}}{a_{\rm max}}\right)^{\frac{1}{2}}-1\right]\,.
\end{align}
The ratio $a_{c,2}/a_{\rm max}$ can be solved from Eq.~\eqref{eq:T} in terms of $T_c/T_{\rm max}$. Using the above equations and the relation $\eta_2=2\eta_{c,2}-\eta_{\rm max}$, one then obtains (for $r\rightarrow 0$)
\begin{align}
\label{eq:a2-over-amax}
    \left(\frac{a_{\rm max}}{a_2}\right)=\left[2\left(\frac{a_{c,2}}{a_{\rm max}}\right)^{\frac{1}{2}}-1\right]^{-2}\,.
\end{align}
For example, for the model discussed in Section~\ref{sup:model}, we have $a_{c,2}/a_{\rm max}\approx 3$ (see Fig.\,\ref{fig:T1overTc}) which gives $a_{\rm max}/a_2\approx 0.16$.
Although we will finally consider a small $a_{\rm max}/a_2$, we keep the dependence on $a_{\rm max}/a_2$ in the following expressions to keep the generality of our analysis.

Before we study how the passage of the wall changes the local radiation energy density, we first take a look at the background radiation energy density outside of the perturbed region, $\bar{\rho}_{\rm rad}(a)$ and $\bar{\rho}_{\rm \chi}(a)$. They are the solution to the following coupled equations
\begin{subequations}
\label{eq:coupled-EoMs}
\begin{align}
    &\frac{\d \bar{\rho}_\chi}{\d t}+3 H \bar{\rho}_\chi = -\Gamma_\chi \bar{\rho}_{\chi}\,,\\
    \label{eq:rad-evolution}
    &\frac{\d \bar{\rho}_{\rm rad}}{\d t}+ 4H \bar{\rho}_{\rm rad}=\Gamma_\chi \bar{\rho}_\chi\,,\\
    & H^2=\frac{1}{3 M^2_{\rm Pl}} (\bar{\rho}_\chi+\bar{\rho}_{\rm rad})\,,
\end{align}    
\end{subequations}
with the initial conditions $\bar{\rho}_{\rm rad}(a_0)=\bar{\rho}_{\rm rad, 0}\approx 0$, $\bar{\rho}_\chi(a_0)=\bar{\rho}_{\chi,0}$ at a certain time $a_0$, determined by the reheating dynamics. These equations are usually solved by assuming $\bar{\rho}_{\rm rad}(a)\ll \bar{\rho}_\chi(a)$, i.e., the energy density is dominated by the reheaton.
The bubble will first generate a perturbed solution for the radiation energy density, which we denote as $\tilde{\rho}_{\rm rad}(a)$ (at the centre of the perturbed region). 

Apparently, for $a< a_1=a_{\rm max}$, we have $\tilde{\rho}_{\rm rad}(a)=\bar{\rho}_{\rm rad}(a)$. At $a=a_1=a_{\rm max}$, part of the radiation energy is transferred to the vacuum energy $|\Delta V_0|$ due to the change of phases. We thus have 
\begin{align}
\tilde{\rho}_{\rm rad}(a_{\rm max})=\bar{\rho}_{\rm rad}(a_{\rm max})-|\Delta V_0|\,.
\end{align}
Between $a_1=a_{\rm max}$ and $a_2$, we can write $\tilde{\rho}_{\rm rad}(a)=\bar{\rho}_{\rm rad}(a)+\delta \rho_{\rm rad}(a)$. In principle, the perturbed solution breaks the homogeneity and one cannot use Eqs.\,\eqref{eq:coupled-EoMs} anymore. But we are going to ignore this inhomogeneity. This way, we have also ignored the diffusion in the generated perturbations. Substituting the said equation into Eq.\,\eqref{eq:rad-evolution}, we obtain
\begin{align}
    \frac{\d \delta\rho_{\rm rad}}{\d t} + 4H \delta\rho_{\rm rad}=0\,,
\end{align}
with the initial condition $\delta\rho_{\rm rad}(a_{\rm max})=-|\Delta V_0|$. We then obtain $\delta\rho_{\rm rad}(a)= (-|\Delta V_0|) (a_{\rm max}/a)^4$ for $a_{\rm max} \leq a< a_2$. The density contrast is then given by $
\delta(a)=(\delta\rho_{\rm rad}(a)+|\Delta V_0|)/\bar{\rho}_{\rm tot}(a)\approx (\delta\rho_{\rm rad}(a)+|\Delta V_0|)/\bar{\rho}_{\chi}(a)$.
At $a_2$, the vacuum energy is transferred back into radiation, and  we have 
\begin{align}
    \label{eq:rho_tilde}
    \tilde{\rho}_{\rm rad}(a_{2})=\bar{\rho}_{\rm rad}(a_{2})-|\Delta V_0| \left(\frac{a_{\rm max}}{a_{2}}\right)^4 + |\Delta V_0|\,.
\end{align}
Similarly, the evolution \emph{after} $a_2$ gives
\begin{align}
    \delta \rho_{\rm rad}(a) &= \left(-|\Delta V_0| \left(\frac{a_{\rm max}}{a_{2}}\right)^4 + |\Delta V_0|\right)\left(\frac{ a_{2}}{a}\right)^4\notag\\ 
    &= |\Delta V_0| \left(\frac{a_{2}}{a}\right)^4
    \left( 
    1 - \left(\frac{a_{\rm max}}{a_{2}}\right)^4
    \right)
\end{align}
where one can replace $a_{\rm max}/a_2$ by Eq.\,\eqref{eq:a2-over-amax}. Dividing the above equation by $\bar{\rho}_\chi (a)$, one may think that the density contrast decreases as $a^{-1}$ after $a_2$. However, so far we have ignored the dynamics of surrounding matter (reheaton $\chi$).
The overdensity in $\rho_{\rm rad}$ can generate a gravitational potential well and accrete the surrounding $\chi$ matter (either particles or an oscillating scalar background field), leading to an overdensity in {\it matter}, $\delta\rho_\chi(a)$. The density contrast is quickly dominated by the contribution from $\delta\rho_\chi$ as the universe is still matter-dominated at this stage. The evolution of $\delta\rho_\chi (a)$ then leads to a linear increase of the total density contrast $\delta(a)$ when $\delta<0.1$~\cite{deJong:2021bbo}. Since the overdensity in radiation $\delta\rho_{\rm rad}$ reaches its maximal value at $a=a_2$, we consider the gravitational effect starting from there. (A more precise description may require a study based on numerical General Relativity, which goes beyond the scope of this work.)
In conclusion, we have 
\begin{align}
\label{eq:delta(a)}
    \delta(a)=\begin{cases}
        0\,, &{\rm for\ } a<a_{\rm max}\\
        \frac{-|\Delta V_0| (a_{\rm max}/a)^4 +|\Delta V_0| }{\bar{\rho}_\chi(a)} \,,  &{\rm for\ } a_{\rm max}\leq a < a_{2}\\
        {\delta_{\rm i} (\frac{a}{a_2})}
        &{\rm for\ } a \geq  a_{2}\,,
    \end{cases}
\end{align}
where
\begin{align}
    \delta_{\rm i} &= \frac{|\Delta V_0|}{\bar{\rho}_\chi(a_{2})} 
    \left( 
    1 - \left(\frac{a_{\rm max}}{a_{2}}\right)^4
    \right)
    \qquad {\rm with \ } a_{\rm i}= a_{2}
    \label{eq:delta-int0}
    \\
    &=\frac{|\Delta V_0|}{\bar{\rho}_{\rm rad}(a_{\rm RH})} 
    \left(\frac{a_2}{a_{\rm RH}}\right)^3
    \left( 
    1 - \left(\frac{a_{\rm max}}{a_2} \right)^4
    \right)\label{eq:delta-int}
    \,,
\end{align}
where in the second line we have used $\bar{\rho}_\chi(a_2)=\bar{\rho}_\chi (a_{\rm RH})(a_{\rm RH}/a_2)^3$ and $\bar{\rho}_{\chi}(a_{\rm RH})=\bar{\rho}_{\rm rad}(a_{\rm RH})$.
Note that a larger $a_{\rm RH}$ leads to a smaller $\delta_{\rm i}$.
This is because the radiation becomes less and less important in the total energy density as we trace back to the past from $a_{\rm RH}$.
On the other hand, $\delta_{\rm i}$ vanishes if $a_{\rm max} = a_2$ corresponding to the case where there is no time for a bubble to grow and shrink.
Since we consider $a_{\rm max}/a_2<1$ not too close to one and its dependence appears with the fourth power, we ignore the $-(a_{\rm max}/a_2)^4$ contribution in the following discussion.

Plugging Eq.\,\eqref{eq:a2-over-amax} into Eq.\,\eqref{eq:delta-int0} with $\bar \rho_\chi (a_2) = \bar \rho_\chi(a_{\rm max}) \cdot (a_{\rm max}/a_2)^3 $, 
$\delta_{\rm i}$ can be re-expressed as
\bal
\delta_{\rm i}
=
\kappa\big(\sqrt{a_2/a_{\rm max}}\big) \frac{|\Delta V_0|}{\bar \rho_\chi (a_{\rm max})},
\eal
where 
\bal
\kappa(y) = 4 y (y-1)
\Big( (2y-1)^4+1 \Big)
\bigg( 1+ \frac{1}{(2y-1)^2} \bigg).
\eal

\subsection{Constraint for successful PBH formation via the post-collapse accretion mechanism}

\begin{figure}
\includegraphics[width=1\linewidth]{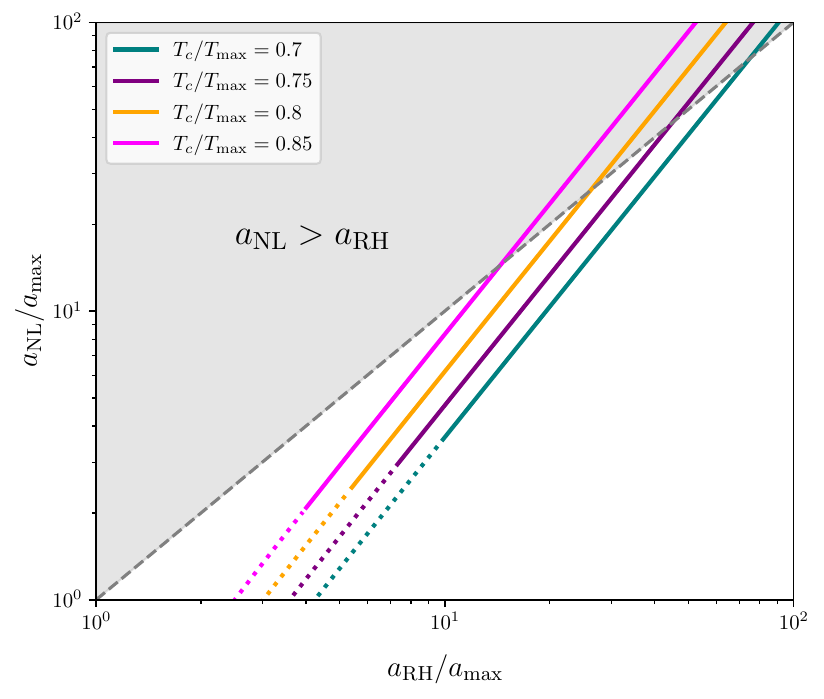}
    \caption{\label{fig:param}
    {Estimation of} $a_{\rm NL}$ as a function of $a_{\rm RH}$, as specified in Eq.~\eqref{eq:NL}. Each plain line corresponds to a different value of the ratio $T_c/T_{\rm max}$, and in this plot, $g_{\star,\phi}(T_c)/g_\star(T_{\rm RH})=0.3$.
    At the bottom of the figure, plain lines turn into dotted lines, as in this regime $a_{\rm NL} < a_{c,2}$, suggesting that the bubble may have already started the collapse into a black hole before the bubble wall turns around.}

\end{figure}

The overdensity $\delta_{\rm i}$ generated at $a= a_{2}$ will induce a gravitational well and accrete the surrounding reheaton into it, leading to the growth of $\delta$. Initially, the growth is linear such that
\begin{align}
    \delta(a)= \delta_{\rm i}\left(\frac{a}{a_{2}}\right)\,.
\end{align}
However, once $\delta(a)$ reaches $\delta_{\rm NL}\sim 0.1$, which defines
\begin{align}
\label{eq:NL}
   a_{\rm NL} &\equiv \left( \frac{0.1}{\delta_{\rm i}}\right) a_2 \,,
\end{align}
the growth becomes non-linear, at an extremely high rate~\cite{deJong:2021bbo}. The non-linear growth quickly leads to the formation of a BH. For this post-collapse accretion mechanism to work in our scenario, we require
\begin{align}\label{eq:criterion}
    a_{\rm NL} < a_{\rm RH}\,.
\end{align}
Substituting Eq.\,\eqref{eq:delta-int} into the above equation, we obtain
\begin{align}
\label{eq:S24}
\frac{\left(\frac{a_{\rm RH}}{a_{\rm max}}\right)^2}{\left[2\left(\frac{a_{c,2}}{a_{\rm max}}\right)^{\frac{1}{2}}-1\right]^4} 
< \frac{10|\Delta V_0|}{\frac{\pi^2}{30}g_\star(T_{\rm RH}) T^4_{\rm RH}} \,.
\end{align}
Assuming a flat potential (which is required to have a large $T_{\rm max}/T_c$), we can estimate $T_c$ as
$|\Delta V_0|\approx (\pi^2/90) g_{\star,\phi} (T_c) T_c^4$. Now from Eq.\,\eqref{eq:T}, we have $(a/a_{\rm max})\approx 4/5^{2/3} (T_{\rm max}/T)^{8/3}$.  Substituting all the relations into Eq.\,\eqref{eq:S24}, we finally obtain
\begin{align}
    \left(\frac{a_{\rm RH}}{a_{\rm max}}\right) <\left(\frac{10 g_{\star,\phi}(T_c)}{3g_\star(T_{\rm RH})}\right)^2 \left(\frac{a_{c,2}}{a_{\rm max}}\right){\left[2-\left(\frac{a_{c,2}}{a_{\rm max}}\right)^{-1/2}\right]^8} \,. 
\end{align}
This gives a constraint on $(a_{\rm RH}/a_{\rm max})$ for a given $(a_{c,2}/a_{\rm max})$.
We can also express the ratios between the cosmological scale factors in terms of the ratios between the temperatures, and obtain
\begin{align}
    &\left( \! \frac{T_{\rm RH}}{T_{\rm max}} \! \right)
    \!\! > \!\frac{8}{5}\left(\frac{3 g_\star(T_{\rm RH})}{10 g_{\star,\phi}(T_c)}\right)^{\!\! 3/4} \! \left[\frac{4}{5^{1/3}} \! \left(\! \frac{T_{\rm max}}{T_c}\! \right)^{\! 1/3} \!\!\! -\left(\! \frac{T_c}{T_{\rm max}} \! \right)\right]^{\! -3} \!\! \!\!.
\end{align}
For example, for $g_{\star,\phi}(T_c)=0.3g_\star(T_{\rm RH}) $, we illustrate the corresponding constraint in Fig.\,\ref{fig:param}, where each coloured line shows $a_{\rm NL}$ in Eq.\,\eqref{eq:NL} as a function of $a_{\rm RH}$ for $T_c/T_{\rm max}=0.7$, $0.75$, $0.8$, and $0.85$. In the grey region, $a_{\rm NL}>a_{\rm RH}$ and PBH formation would not take place since the matter domination finishes before starting the non-linear collapse. However, in the lower region, $a_{\rm NL}<a_{\rm RH}$ and the collapse can take place before the onset of radiation domination. The dotted region at the bottom of each line corresponds to $a_{\rm NL} < a_{c,2}$, suggesting that the collapse should already start before the bubble wall turns around. In such a case, the PBH formation still takes place, although our equations in this appendix may not be valid.

\bibliographystyle{apsrev4-1}
\bibliography{Ref}
\end{document}